\documentclass[twocolumn,pra,aps,superscriptaddress,showpacs]{revtex4}
\usepackage{mathrsfs}
\usepackage{graphicx}
\begin{document}
\title {Disorder effects on the quantum coherence of a many-boson system}
\author{Zhi-Hai Zhang}
\author{Pei Lu}
\author{Shiping Feng}
\author{Shi-Jie Yang\footnote{Corresponding author: yangshijie@tsinghua.org.cn}}
\affiliation{Department of Physics, Beijing Normal University,
Beijing 100875, China}
\begin{abstract}
The effects of disorders on the quantum coherence for many-bosons
are studied in a double well model. For the ground state, the
disorder enhances the quantum coherence. In the deep Mott regime,
dynamical evolution reveals periodical collapses and revivals of the
quantum coherence which is robust against the disorder. The average
over variations in both the on-site energy and the interaction
reveals a beat phenomenon of the coherence-decoherence oscillation
in the temporal evolution.

\end{abstract}
 \maketitle

The ultracold atoms in an optical lattice have provided an
unprecedent opportunity to explore the correlated many-body problems
and nonlinear dynamics in condensed matter
physics\cite{Will,Jaksch,Chin,Bloch,Greiner,Sebby}. The atoms
tunneling between lattice sites play the role of electrons in the
solid. Due to the controllable manipulation of the optical lattice,
the inter-particle interaction as well as the well developed
measurement technique, scientists can now precisely examine the
coincidence of theoretical models, e.g., the Hubbard model in
strongly correlated systems, with
experiments\cite{Sciolla,Wolf,Tiesinga}. On the other hand,
experimental techniques also exploited the collapse and revival
dynamics of ultracold atoms in the optical
lattice\cite{Bloch2,Morsch}. In a recent experiment, S. Will et al
observed the beat phenomenon of quantum coherence for Bose atoms in
the dynamical evolution\cite{Will}. This phenomenon was interpreted
as the multi-body interactions due to multi-orbital virtual
transitions.

The disorders usually play important roles in realistic materials.
It profoundly affects the transport properties of a wide range of
materials. The dilute atomic gas in disordered optical potentials is
again an ideal candidate model to simulate various disorder effects
in condensed matter physics\cite{Lye,Fort,Schulte,Fallani}.
Disorders may be created in the laser potential in different ways.
One of them is to add a speckle pattern to the regular
lattice\cite{White}. The disorder strength is continuously tunable
by controlling the intensity of the speckle field. The localization
effects due to correlation and disorder compete against each other,
resulting in a partial delocalization of the particles in the Mott
regime, which in turn lead to increased phase
coherence\cite{Sengupta}. However, there is significant disagreement
regarding features of the disordered Bose-Hubbard
model\cite{Lewenstein,Sanchez,Pollet,Altman,Zhou,Deissler}.
Experiments are still some way from the appropriate regime to
observe, e.g., Anderson localization, due to the large speckle size
and the delocalizing effect of the interatomic interactions. The
interplay between the interaction and the disorder also lead to a
rich and complex arrangement of different insulating and superfluid
states\cite{Fisher,Giorgini,Gimperlein}.

In this paper, we explore the quantum coherence of a correlated Bose
gas in a double well system\cite{Anderson,Cataliotti,Albiez}. We
show that the on-site energy disorder enhance the quantum coherence
in the ground state. In the deep Mott regime, dynamical evolution
reveals periodical collapses and revivals of the quantum coherence.
The periodicity is robust against the disorder. In addition, the
disorder in the interatomic interaction also plays an important
role. We find that the revival oscillations of quantum coherence
exhibit the beat phenomenon when a moderate interaction disorder is
taken accounted of. The corresponding Fourier spectra of the
coherence parameter exhibit a double-peak structure.

We are concerned with $N$ bosonic atoms confined in a double well
potential. By considering only the lowest energy band, the
Bose-Hubbard Hamiltonian is written as \cite{Vardi,Yang}
\begin{equation}
\hat{H}=-t(\hat{a}_1^\dag\hat{a}_2+\hat{a}_2^\dag\hat{a}_1)+\frac{U}{2}\sum_i
\hat{n}_i(\hat{n}_i-1)+\varepsilon (\hat{n}_1-\hat{n}_2),
\end{equation}
where $\hat{a}_i^\dag$ and $\hat{a}_i$ ($i=1,2$) are the creation
and annihilation operators in either side of the well. $\hat{n}_i$
are the number operators. $U$ and $t$ are the Hubbard energy and the
atom hopping, respectively. $\varepsilon$ is the on-site energy
difference between the two sites. The disorder is introduced into
the model via the fluctuation of $\varepsilon$ , which is assumed to
randomly and uniformly distribute within the interval
$-\bigtriangleup\varepsilon\leq \varepsilon \leq
\bigtriangleup\varepsilon$.

The Hamiltonian $(1)$ is explicitly represented in the Fock basis
set $\{|N,0\rangle, |N-1,1\rangle, \cdots, |0,N\rangle \}$. The
general eigenstates are expressed as linear combinations of the
occupation bases, $|\psi_j\rangle=\sum_{k=0}^N c_{jk}|N-k,k\rangle$
($j=0,1,2,\ldots,N$), which correspond to the eigenvalues
$\omega_j$. The coefficients $c_{jk}$ satisfy the recursive relation
\begin{widetext}
\begin{equation}
-t\sqrt{(N-k)(k+1)} c_{j(k+1)}-t\sqrt{(N-k+1)k} c_{j(k-1)}+
[\frac{U}{2}(N^2-2Nk-N+2k^2)+\varepsilon (N-2k)-\omega_j] c_{jk}=0.
\end{equation}
\end{widetext}
The temporal evolution of the state is governed by the
Schr\"{o}dinger equation for a given initial state
$|\psi(0)\rangle$:
\begin{equation}
|\psi(\tau)\rangle=\sum_{j=0}^Nf_j(\tau)|\psi_j\rangle= \sum_{k=0}^N
g_k(\tau) |N-k,k\rangle ,
\end{equation}
where $g_k(\tau)=\sum_{j=0}^N f_j(0) c_{jk} e^{-i\omega_j \tau}$,
with $f_i(0)=\langle\psi_j|\psi(0)\rangle$.

To depict the coherence degree of the system, we introduce a
characteristic parameter\cite{Yang}:
\begin{equation}
\alpha(\tau)=\frac{\lambda_1-\lambda_2}{\lambda_1+\lambda_2}
\end{equation}
where $\lambda_1$ and $\lambda_2$ are the two eigenvalues of the
single-particle density $\rho_{\mu\nu}(\tau)=\langle
\psi(\tau)|\hat{a}_\mu^\dag \hat{a}_\nu |\psi(\tau)\rangle$
($\mu,\nu=1,2$)\cite{Penrose,Mueller}. When $\alpha\rightarrow 1$,
the system is in the coherent (quasicoherent) state since in this
case there is only one large eigenvalue of the matrix
$\rho_{\mu\nu}$. Accordingly, $\alpha\rightarrow 0$ indicates the
system is in the decoherent or fragmented state because there are
two densely populated natural orbits. In the weak-interaction or
strong-tunneling limit ($U/t\ll 1$), each atom is in a coherent
superposition of the left-well and the right-well states. The ground
state of the system is a state with a mean number $N/2$ of atoms in
each well. In the opposite limit of the strong-interaction or weak
tunneling ($U/t\gg 1$), the tunneling term is negligible. In this
case, the Hamiltonian is the product of number operators for the
left and right wells. The eigenstates are products of Fock states
and are referred as decoherent states. This regime is analogous to
the Mott insulator (MI) phase in a lattice system.

In the following we consider the typical cases in which the initial
state is a coherent state. We choose $N=10$ and set $t=1$ as the
energy units. All results are taken $50$ repetitions of the random
choice of disorders, which is enough to average out the
fluctuations.

\begin{figure}
\begin{center}
\includegraphics*[width=10cm]{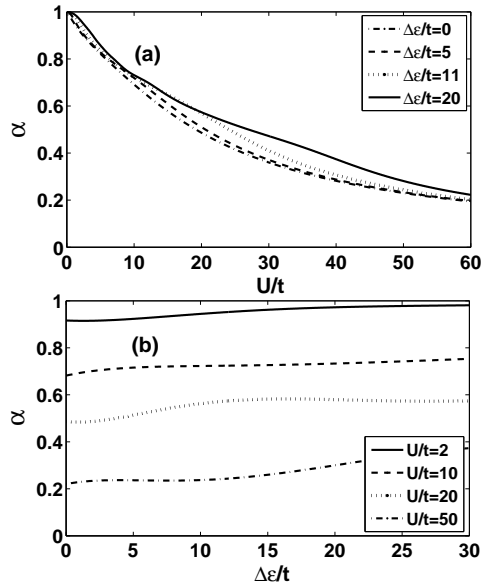}
\caption{(a) $\alpha$ versus $U/t$ for several on-site disorder
strength (from lower to upper) $\Delta \varepsilon/t=0, 5, 11, 20$.
(b) $\alpha$ versus $\Delta \varepsilon/t$ for several values of
Hubbard energies. From upper to lower: $U/t=2, 10, 20, 50$.}
\end{center}
\end{figure}

Figures 1 display the coherence degree $\alpha$ versus the disorder
in the ground states by taking average over the on-site energy
within the range of $\Delta \varepsilon/t=0,5,11,20$, respectively.
From Fig.1 (a), it can be seen that in all range of $U/t$ the
disorder raises $\alpha$, in agreement with previous studies. The
coherence increase by the disorder is most prominent in the regime
of intermediate strength of the Hubbard interactions. Fig.1 (b)
shows the disorder dependence of the coherence degree $\alpha$ at
several values of the Hubbard energies $U/t=2,10,20,50$,
respectively. For smaller $U/t$, the ground state has larger
$\alpha$, indicating that it is in a coherent (superfluid) state.
For all $U$, the on-site energy disorder helps to enhance the
quantum coherence. Generally, the stronger disorder has the larger
coherence degree.
\begin{figure}
\begin{center}
\includegraphics*[width=10cm]{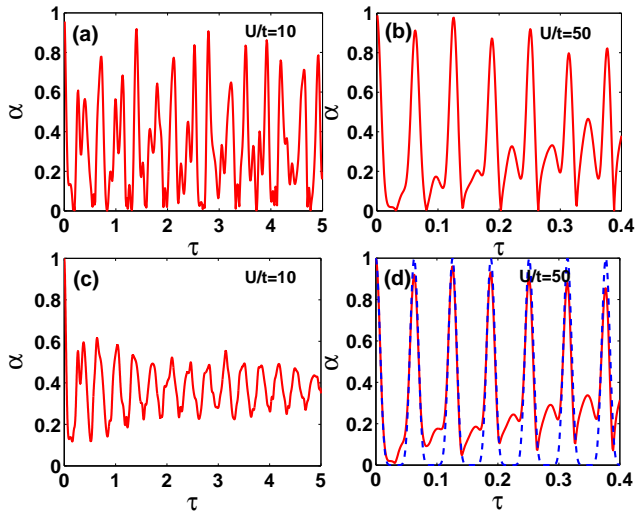}
\caption{(color online) The temporal evolution of the coherence
degree $\alpha(\tau)$ without the disorder for (a) $U/t=10$ and (b)
$U/t=50$, and with the on-site disorder $\Delta \varepsilon/U=20\% $
for (c) $U/t=10$ and (d) $U/t=50$. The dashed line in (d) is for the
independent evolution as explained in the text.}
\end{center}
\end{figure}

Next we study the disorder effect on the quantum dynamics of the
$N$-bosons system. Suppose the system is initially in a coherent
state,
\begin{equation}
|\psi(0)\rangle=(\frac{\hat{a}_1^\dag+\hat{a}_2^\dag}{\sqrt{2}})^N
|0\rangle.
\end{equation}
The lattice parameter is non-adiabatically changed to the deep MI
regime. Figures 2 are the temporal evolution of the coherence degree
$\alpha$ for two typical interactions $U/t=10,50$, respectively. It
is clearly shown that $\alpha$ oscillates between one and zero,
implying the system experiences revivals and collapses of the
coherence. Fig.2 (a) and (b) are for the system without the
disorder. Clearly, the periodicity is more distinct for larger
$U/t$. The period of the oscillation is $T=\pi/U$. Fig.2(c) and (d)
present temporal evolution of $\alpha$ with the on-site disorder
$\Delta \varepsilon/U=20\% $ for two different values of $U/t=10$
and $U/t=50$, respectively. For small $U/t$ (Fig.2(c)), the
amplitude of the oscillation is small. The large amplitude of
periodical oscillation of $\alpha(\tau)$ in Fig.2(d) implies the
system experiences revivals and collapses of the coherence. In
contrary to one's intuition that the disorder may destroy the
periodicity of the revival and collapse oscillation, they enhance
the periodicity by damping the irregular fluctuations of the
coherence degree $\alpha$ in the temporal evolution.

This phenomenon can be understood as follows. The initial state can
be expanded as $|\psi(0)\rangle \sim \sum_{k=0}^N C_N^k
(\hat{a}_1^\dag)^{N-k} (\hat{a}_2^\dag)^k |0\rangle$. When $U/t\gg
1$, the Fock states $(\hat{a}_j^\dag)^k |0\rangle$ in each well are
the eigenstates, and the corresponding eigenenergies are
$Uk(k-1)/2$. The system is then a superposition of products of the
Fock states, which evolves independently as,
\begin{widetext}
\begin{eqnarray}
|\psi(\tau)\rangle &\sim& \sum_{k=0}^N C_N^k
e^{-i[\frac{1}{2}U(N-k)(N-k-1)+\varepsilon_1 (N-k)]\tau}\times
(\hat{a}_1^\dag)^{(N-k)}e^{-i[\frac{1}{2}Uk(k-1)+\varepsilon_2
k]\tau}(\hat{a}_2^\dag)^k
|0\rangle \nonumber\\
&=& e^{-i[\frac{1}{2}UN(N-1)+\varepsilon_1 N]\tau} \sum_{k=0}^N
C_N^k e^{i[U(N-k)k+\varepsilon k]\tau}
(\hat{a}_1^\dag)^{N-k}(\hat{a}_2^\dag)^{k}|0\rangle,\label{independent}
\end{eqnarray}
\end{widetext}
where a time-dependent phase factor is attached in each term. If
there are no disorders ($\Delta \varepsilon=0$), the system will
recover its initial state when the time evolves an integer multiples
of $\pi/U$. In between this period, superposition form various terms
cancels and the coherence is destroyed\cite{Yang}. In the presence
of the disorder, there are additional phases associating to the
$\Delta \varepsilon$ in each term that seem destroy the coherence of
the quantum state. However, numerical calculations show that it is
not true. The origin is that in the deep Mott regimes, particle
tunneling is suppressed. All Fock states have almost the same
oscillating periods which only depends on the interaction strength
$T=\pi/U$. Hence average over onsite energy $\epsilon$ does not
reduce the amplitude. The dashed line in Fig.2(d) reveals that the
periodicity are perfectly reserved. We conclude that the phenomenon
of the revivals and collapses of the coherence is robust against the
disorder\cite{Greiner}.

\begin{figure}
\begin{center}
\includegraphics*[width=10cm]{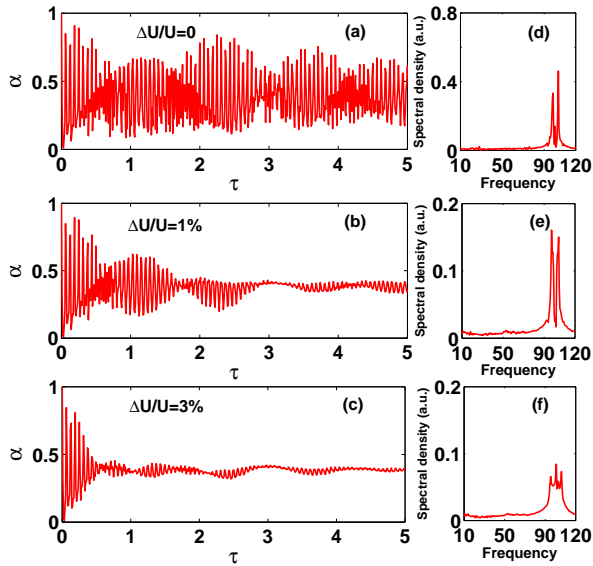}
\caption{(color online) Left panels: Temporal evolution of the
coherence degree $\alpha$ with both on-site disorders
$\Delta\varepsilon/U=20\%$ and interaction disorders. (a) $\Delta
U/U=0$, (b) $\Delta U/U=1\%$, and (c) $\Delta U/U=3\%$, with
$U/t=50$. Right panels: the corresponding Fourier spectra.}
\end{center}
\end{figure}

We explore the system by further taking account of the interaction
disorders with the Hubbard energy $U$ in a rang of uniformly
distributed interval $U\pm\Delta U$. The on-site disorder is fixed
at $\Delta\varepsilon/U=20\%$ with $U/t=50$. Figures 3 show the
temporal evolution of $\alpha$ for various interaction disorders
(left panels). The right panels (d-f) are the corresponding Fourier
spectra which exhibit double-peak structures as signals of beat
effects. In the absence of the interaction disorders ($\Delta
U/U=0$), Fig.3(a) shows that the beat effect is smeared by
additional oscillations in the temporal evolution of $\alpha$, which
is evidently shown in Fig.2(b). These oscillations give rise to an
additional peak at two times of the double-peak frequency. When
interaction disorders are counted in (Fig.3(b) and (c)), the
additional oscillations in Fig.2(b) are damped, as shown in
Fig.2(d). Consequently, the interaction disorder enhances the beat
effect of the coherence-decoherence osccilations.
\begin{figure}
\begin{center}
\includegraphics*[width=11cm]{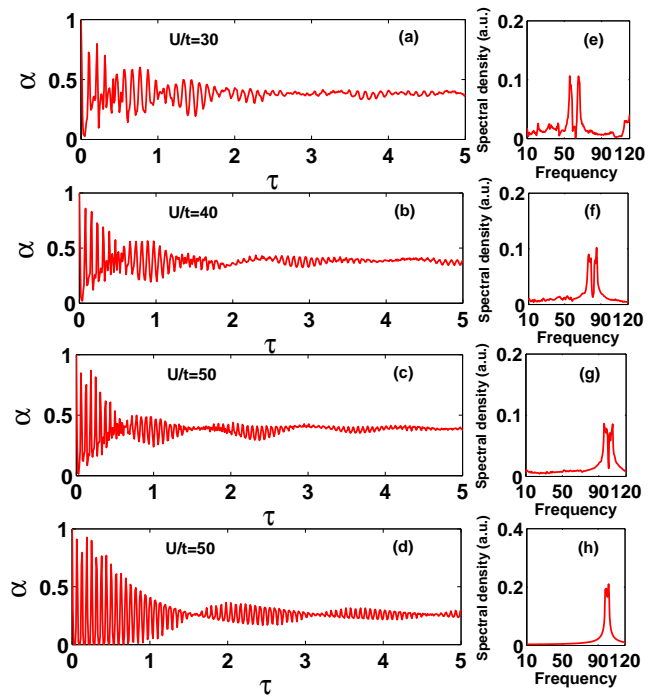}
\caption{(color online) Left panels: Temporal evolution of the
coherence degree $\alpha(\tau)$ with both on-site disorders
$\Delta\varepsilon/U=20\%$ and interaction disorders $\Delta
U/U=2\%$ for (a) $U/t=30$, (b) $U/t=40$, and (c) $U/t=50$. For
comparison, (d) is independent evolution (\ref{independent}) with
both disorders. Right panels: the corresponding Fourier spectra.}
\end{center}
\end{figure}

Figures 4 show the temporal evolution of $\alpha$ for various
interactions $U/t=30, 40, 50$, respectively, with
$\Delta\varepsilon/U=20\%$ and $\Delta U/U=2\%$. As a comparison,
Fig.4(d) displays the result from independent evolution described by
equation (\ref{independent}). Figs.4 (e)-(h) are the corresponding
Fourier spectra. It shows that in deeper MI regime, the beat effects
become more prominent.

In a recent paper, it has been observed the beat phenomenon of the
revivals and collapses of the quantum coherence in an optical
lattice in a longer time interval\cite{Will}. The beat signal of the
revival and collapse oscillation was attributed to the multi-body
interactions, emerging through virtual transitions of particles from
the lowest energy band to the higher energy bands. In our model, the
atoms are assumed to occupy a single spatial orbital and only the
two-body interaction Hubbard energy $U$ is independent of the
filling at the lattice site. The disorder can enhance the
periodicity of the revival and collapse oscillation by damping the
higher-order frequencies. It is not clear yet if there are inherent
relations between the beat phenomena observed by S. Will et al and
our theoretical model.

In summary, we have investigated the disorder effect on the quantum
coherence of a many-boson system. We find that the disorder enhance
quantum coherence in the ground states. Dynamical evolution of the
cold Bose atoms exhibits collapses and revivals of the coherence
which is robust against the disorder. In particular, the interaction
disorders generate the beat phenomenon of coherence in the temporal
evolution. It may provide a signal to observe the effects of the
interaction disorders experimentally.

This work is supported by the funds from the Ministry of Science and
Technology of China under Grant Nos. 2012CB821403 and 2011CB921700
and by the National Natural Science Foundation of China under grant
No. 10874018.

\end{document}